# Graphene Heat Spreaders for Thermal Management of Nanoelectronic Circuits

Samia Subrina, Dmitri Kotchetkov and Alexander A. Balandin

*Abstract* — **Graphene was recently proposed as a material for heat removal owing to its extremely high thermal conductivity. We simulated heat propagation in silicon-on-insulator (SOI) circuits with and without graphene lateral heat spreaders. Numerical solutions of the heat propagation equations were obtained using the finite element method. The analysis was focused on the prototype SOI circuits with the metal – oxide - semiconductor field-effect transistors. It was found that the incorporation of graphene or few-layer graphene (FLG) layers with proper heat sinks can substantially lower the temperature of the localized hot spots. The maximum temperature in the transistor channels was studied as function of graphene's thermal conductivity and the thickness of FLG. The developed model and obtained results are important for the design of graphene heat spreaders and interconnects.**

## I. INTRODUCTION

Over many years silicon technology demonstrated continuous improvement in both performance and productivity. Silicon-on-insulator (SOI) wafers and designs offer major advantages over traditional silicon device structures such as improved electrical isolation, reduced parasitic capacitances, improved radiation hardness and higher packing density. At the same time, the buried oxide in SOI metal-oxide-semiconductor field-effect transistor (MOSFET) structures insulates the active channel from the substrate not only electrically but also thermally. As a result, the temperature rise in SOI MOSFETs can become excessive leading to performance degradation and early thermal breakdowns [1] – [3]. The down-scaling and higher circuit speeds lead to further increase in heat generation, power densities and temperature rise [4] – [5]. Efficient thermal management becomes an integral part of the device design for long-term reliability and optimum performance.

This work was supported by DARPA—SRC Focus Center Research Program (FCRP) through its Center on Functional Engineered Nano Architectonics (FENA) and Interconnect Focus Center (IFC) and by AFOSR Award A9550-08-1-0100 on Phonon Engineered Heterostructures.

Samia Subrina, Dmitri Kotchetkov, Alexander A. Balandin (corresponding author; e-mail: balandin@ee.ucr.edu) are with the Department of Electrical Engineering, University of California, Riverside, CA 92521 USA (see web-site at: www.ndl.ee.ucr.edu).

One possible solution for removing heat from the localized hot spots is to incorporate chips with materials that have very high thermal conductivity, i.e. high-heat flux (HHF) thermal management.

Graphene, the latest of the discovered allotropes of carbon, exhibits extremely high intrinsic thermal conductivity. The room temperature (RT) thermal conductivity of the suspended graphene was determined to be ranging from 3080 - 5300 W/mK [6] – [7]. It also revealed strong flake size dependence [7] – [9]. Many studies demonstrated that quality graphene has very high electron mobility (~$2\times10^5$ $cm^2V^{-1}s^{-1}$) and low resistivity (~$10^{-8}$ Ohm-m) [10] – [13]. The unique electrical and thermal properties of graphene make it a promising material for the low-noise transistors [14], electrical interconnects [15] and thermal management [6] - [9]. In this work we carry out a feasibility study of graphene applications as lateral heat spreaders.

## II. MODEL FOR HEAT CONDUCTION

To evaluate the feasibility of the use of graphene for thermal management, we simulate heat propagation in SOI structures with and without graphene layers. In our model we approximate several MOSFETs as rectangular channels, separated from each other by 10 μm, which generate heat (see Fig. 1). The thicknesses of Si substrate, buried oxide layer and surface Si film are 500 μm, 100 nm and 25 nm, respectively, while their thermal conductivities are 155 $Wm^{-1}K^{-1}$, 1.38 $Wm^{-1}K^{-1}$ and 155 $Wm^{-1}K^{-1}$, respectively [16]. A conventional heat sink is attached to the device structure at the bottom. A graphene heat spreader layer, when used, is sandwiched between the oxide layer and Si substrate. Its two ends are attached to the side heat sinks (or otherwise connected to the bottom sink), thus forming the channel for heat escape.

The simulations were carried out with the finite element method (FEM) using COMSOL software [17]. The heat transport along graphene layers was assumed to be diffusive since the size of the layer is much larger than the phonon mean free path [6] – [9]. The heat conduction was modeled by solving numerically Fourier's law

$$-\nabla\cdot(k\nabla T)=Q, \qquad (1)$$

where $Q$ is the heat source, defined as the heat energy generated within a unit volume per unit time, $T$ is the absolute temperature and $k$ is the thermal conductivity. The bottom surface of the substrate and the two opposite ends of the graphene heat spreader were kept at a constant temperature $T_0$ = 300 K. The external surfaces were modeled as insulated from environment.

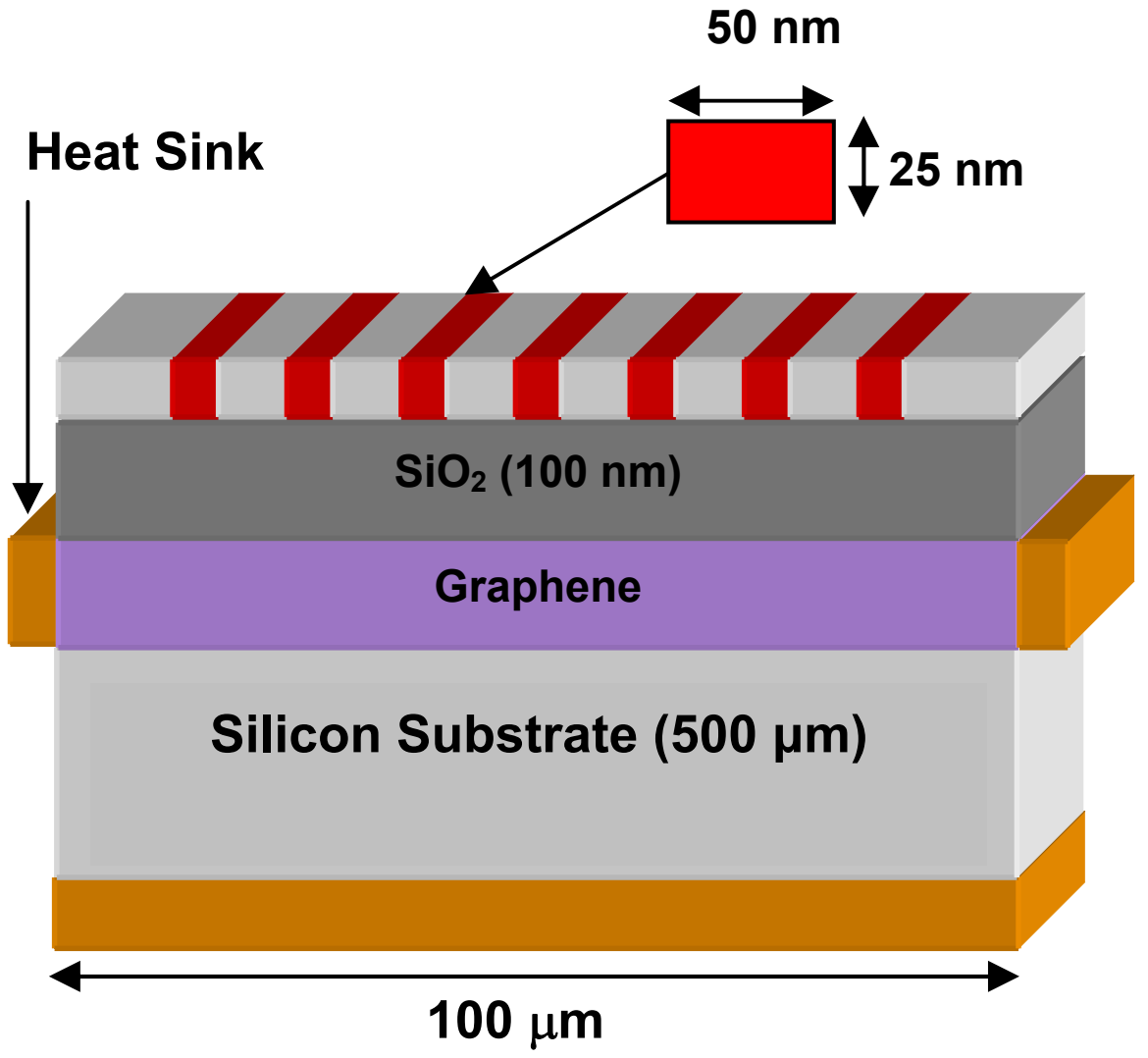

Fig. 1: Schematic of a circuit on SOI substrate with the graphene lateral heat spreader attached to side heat sinks. The thicknesses are not to scale.

## III. SIMULATION RESULTS

In the simulations, the thermal conductivity of graphene or FLG was assumed to be ranging from 1000 $Wm^{-1}K^{-1}$ to 5000 $Wm^{-1}K^{-1}$ [8]. The decrease in the value of graphene thermal conductivity from the maximum reported for the large suspended flakes [6] can come up as a result of the flake size dependence, temperature rise and the interface contact effect [8-9]. Fig. 2 shows the calculated temperature profiles for the SOI circuit (a) without and (b) with the graphene lateral heat spreader. The temperature color scheme is indicated in the sidebars. The linear power density of each active channel was set to 0.5 W/mm. For given parameters of the structure and a number of transistors, the maximum temperature in the hot spots decreases by 70 K when graphene layers are embedded in the chip. The effect of the graphene lateral heat spreaders is more pronounced when the number of transistors increases.

In Fig. 3 we show a comparison between the temperature rise in a SOI chip with one MOSFET and a chip with seven MOSFETs. The temperature drop owing to incorporation of the single-layer graphene heat spreader is about ~ 23 % in the seven finger chip while in the structure with one device it is ~ 11 %. The overall cooling of the device structures with the lateral heat spreaders depends on the distance between the heat generating devices, their geometry, thickness of the substrate, overall size of the chip and power dissipated in each device.

It is important to understand how the heat spreading ability of graphene or FLG layer depends on its thickness and the value of the thermal conductivity. The chip cooling with graphene under-layers is within the general approach of HHF thermal management. It might be more technologically feasible for HHF heat removal to use FLG with larger cross-sectional area instead of single layer graphene. The thermal conductivity of FLG is expected to be lower than that of graphene approaching that of bulk graphite (around ~2000 $Wm^{-1}K^{-1}$ at RT along the basal plane). The mechanical and thermal properties of FLG are less subject to degradation when the layer is embedded between the oxide and the substrate.

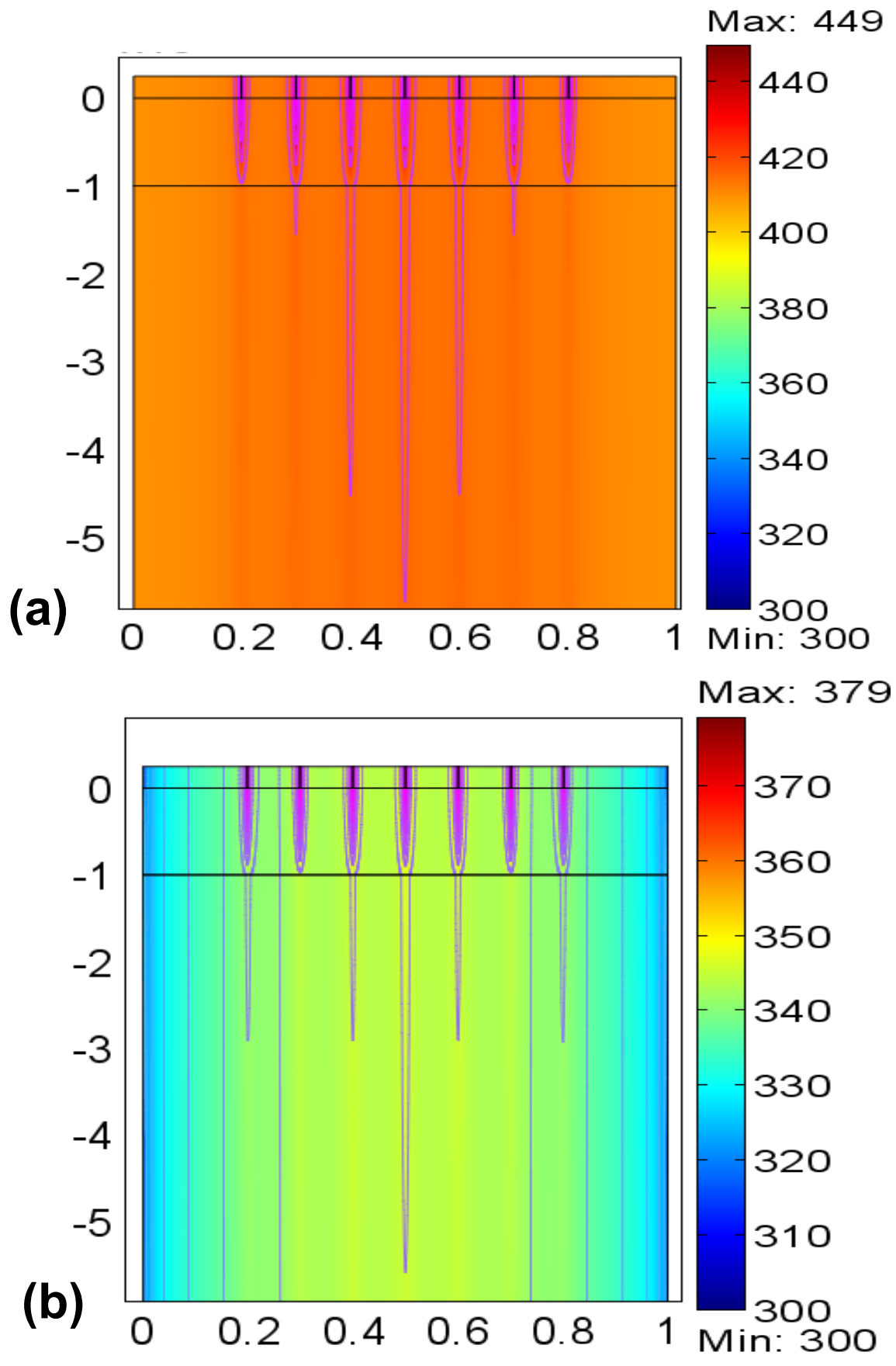

Fig. 2: Temperature distribution across SOI-based circuit with seven active transistors (a) without and (b) with graphene heat spreaders attached to the heat sinks. The spreaders are embedded between $SiO_2$ layer and the substrate. The thermal conductivity of graphene is assumed to be 5000 $Wm^{-1}K^{-1}$.

Fig. 4 shows the simulated maximum temperature of SOI chip as a function of the thermal conductivity value and thicknesses of the heat spreader. The general trend is that the cooling becomes more efficient with the increasing thermal conductivity and thickness of the spreader owing to increasing heat flux directed away from the hot spots. Although the thermal conductivity of FLG has not been reported yet, one can expect it to be between the values measured for single layer graphene [6] and bulk graphite. Our results suggest that the lateral heat spreaders with the number of atomic planes between 3 and 10 would be efficient. This is encouraging news for practical applications because it allows for greater flexibility with fabrication.

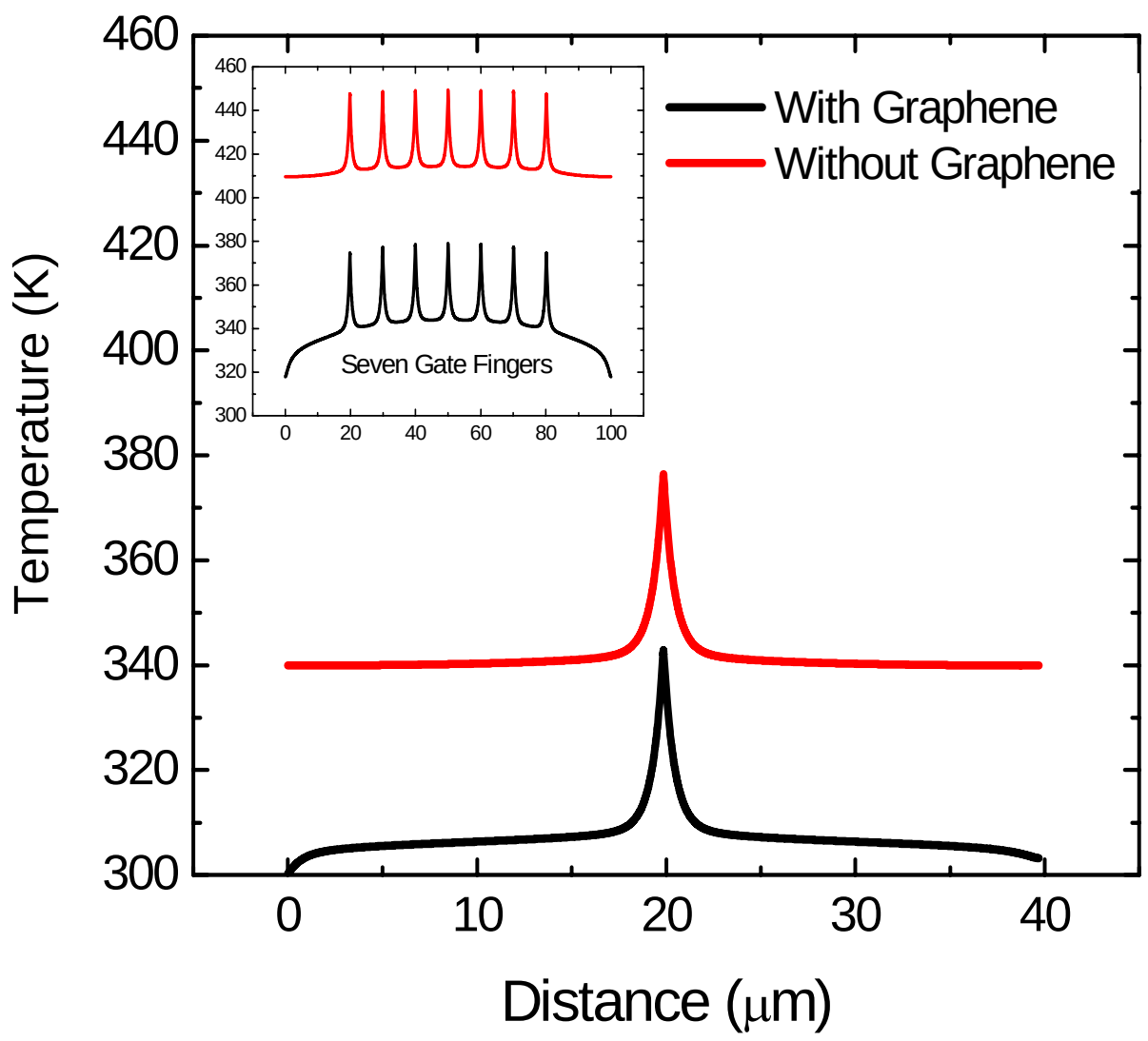

Fig. 3: Temperature profile along the top surface of a SOI-based MOSFET with (black) and without (red) graphene heat spreaders. The inset shows the temperature profile for a SOI chip with seven active devices (fingers).

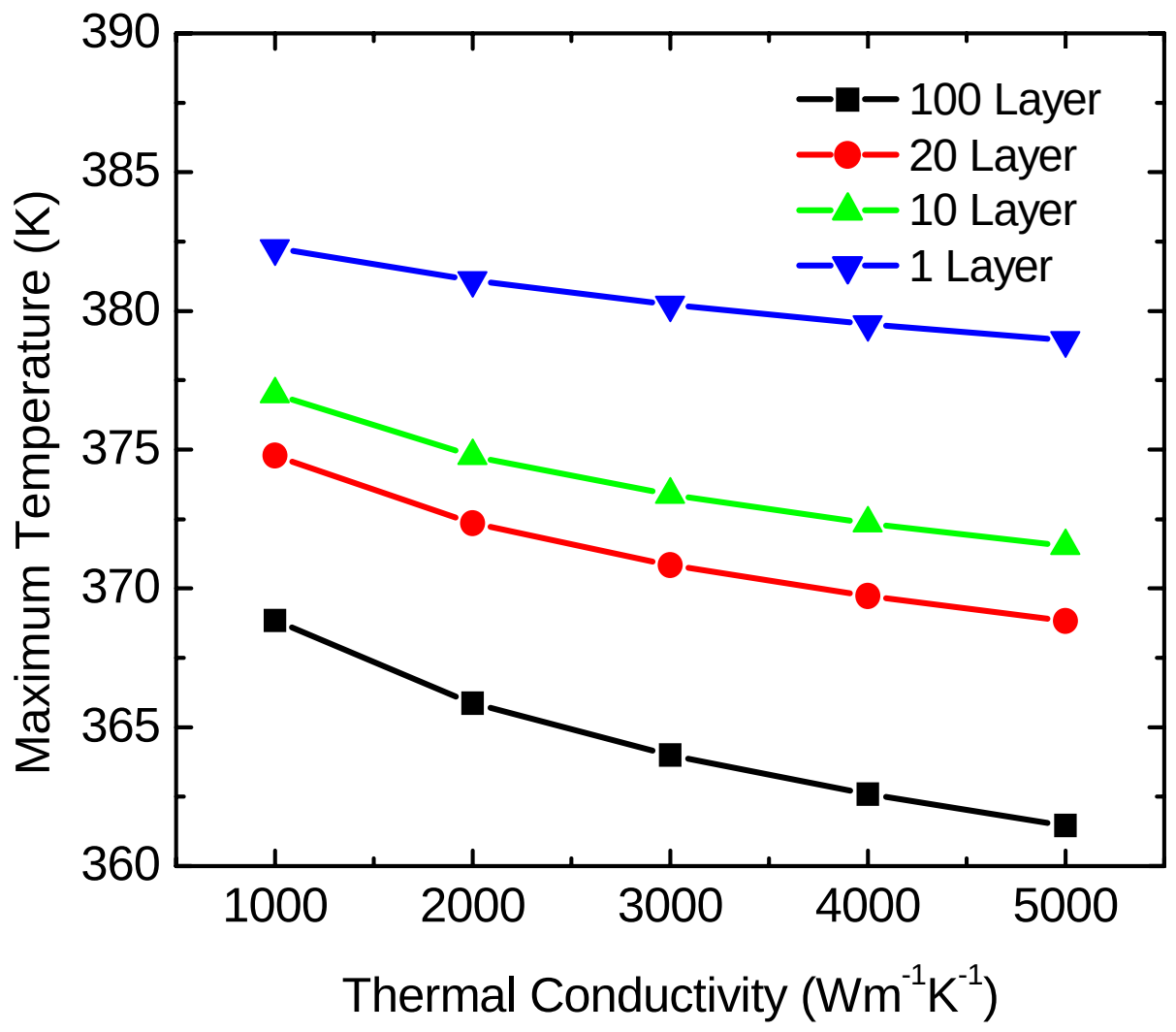

Fig. 4: Maximum temperature in the SOI chip as a function of the thermal conductivity of the lateral heat spreader. The number of graphene atomic planes in the layers was varied from 1 to 100.

Continuing progress in epitaxial, chemical and other synthesis techniques of graphene [18-19] may allow for large-area growth of the lateral FLG heat spreaders. The epitaxial graphene is less subject to possible adhesion problems for mechanically exfoliated graphene. Although the high-quality low-noise graphene transistors with the top gate, where the exfoliated graphene was embedded between two electrically insulating oxide layers, have already been demonstrated [20]. The thermal interface contact issue for graphene with the heat sinks is much less severe than that for carbon nanotubes owing to larger overlap area [21]. At the same time, graphene – $SiO_2$ interfacial thermal contact resistance may produce deleterious effects on the hot remediation even if the lateral thermal conductivity is extremely high [22]. This issue requires further study in the context of graphene heat spreaders. HHF thermal management with graphene or FLG may work even better for 3-D circuits [23].

## IV. SUMMARY

We carried out a feasibility study of the use of graphene as the material for lateral heat spreaders in SOI based chips. Our results show that the incorporation of graphene or FLG under the insulating layer can lead to substantial reduction in temperature of the hot spots. The efficiency of the hot spot removal with graphene depends on the specifics of the device structure and geometry. Numerical experiments suggest that FLG heat spreaders may be technologically more feasible than single layer graphene. The obtained results can lead to a new type of HHF thermal management of nanoelectronic circuits.